\documentclass[prl,twocolumn,superscriptaddress]{revtex4-1}
\usepackage{graphics,graphicx,natbib}
\newcommand{\eq}{&=&}
\newcommand{\eqn}[1]{\begin{eqnarray*}#1\end{eqnarray*}}
\newcommand{\eqnn}[1]{\begin{eqnarray}#1\end{eqnarray}}

\begin{document}

\title{Entangling moving cavities in non-inertial frames}

\author{T. G. Downes}
\email{downes@physics.uq.edu.au}
\affiliation{Centre for Quantum Computer Technology, Department of Physics, The University of Queensland, Brisbane 4072 Australia}
\author{I. Fuentes\footnote{Published before under Fuentes-Guridi and Fuentes-Schuller}} 
\affiliation{School of Mathematical Sciences, University of Nottingham, Nottingham NG7 2RD, United Kingdom}
\affiliation{Institute for Theoretical Physics, Technical University of Berlin, Hardenbergstr. 36, D-10623, Berlin, Germany}
\author{T. C. Ralph}
\affiliation{Centre for Quantum Computer Technology, Department of Physics, The University of Queensland, Brisbane 4072 Australia}

\begin{abstract}
An open question in the field of relativistic quantum information is how parties in arbitrary motion may distribute and store quantum entanglement. We propose a scheme for storing quantum information in the  field modes of cavities moving in flat space-time and analyze it in a quantum field theoretical framework. In contrast to previous work that found entanglement degradation between observers moving with uniform acceleration, we find the quantum information in such systems is protected. We further discuss a method for establishing the entanglement in the first place and show that in principle it is always possible to produce maximally entangled states between the cavities.
\end{abstract}

\maketitle

In the emerging field of relativistic quantum information, space-time and relativistic effects are incorporated into the question of how to process information using quantum systems.  The fact that nature is both quantum and relativistic, and that many realistic implementations of quantum information involve relativistic systems, has motivated the development of this research field \cite{peres:2004}. A key step is to find suitable ways to distribute and store quantum entanglement in relativistic scenarios. Entanglement is an essential resource for most quantum information protocols \cite{nielsen:2000}. Proposals need to be studied in a fully quantum field theoretical framework. In this way we can find rigorous theoretical upper bounds on the quality of entanglement that can be shared between observers in arbitrary motion.

In this paper we show that, within the framework of quantum field theory, there is no upper bound on the entanglement that can be shared between an inertial and a uniformly accelerating observer. This is in contrast to previous studies on the effects of uniform acceleration on entanglement which show that the entanglement is degraded from the perspective of accelerating observers. These studies include entanglement between free fields \cite{fuentes:2005, *adesso:2007, *pan:2008} and entanglement between atoms \cite{lin:2008,*landulfo:2009}. This degradation of entanglement is usually attributed to the Unruh effect, where a uniformly accelerating observer will perceive the Minkowski vacuum as being populated with a thermal distribution of particles \cite{unruh:1976}. Alsing and Milburn  \cite{alsing:2003}  introduced the idea of using moving cavities to store quantum information and concluded that  any attempt to implement a teleportation protocol in non-inertial frames would be hampered by the presence of Unruh radiation. Unfortunately, only free field modes were explicitly used in their calculation as pointed out  by Schutzhold and Unruh \cite{Schutzhold:2005p311}.  

Here we reconsider the idea of using moving cavities in space-time to store quantum information including the boundary conditions necessary to describe the field inside the cavities.  When such boundary conditions are taken into account, it becomes clear that the cavity walls protect the entanglement once it has been created.  Furthermore, we show that entangling an inertial and a non-inertial cavity is non-trivial but that it can always be achieved in principle. Thus we demonstrate that there is no theoretical upper bound to the quality of the entanglement that can be shared between an inertial and uniformly accelerated observer. Throughout this paper, we work in natural units $c=\hbar=1$.

We consider  an observer, called Alice,  stationary in Minkowski coordinates $(t,x)$ who holds a cavity.  Alice will encode quantum information in a massless scalar field contained within the cavity's walls which are  described by two mirrors, one at $x_1$ and the other at $x_2$ with $\left|x_2-x_1\right|=L$.

\setlength{\unitlength}{1cm}

\begin{figure}[hc]
\begin{center}
\includegraphics[width=6.5cm]{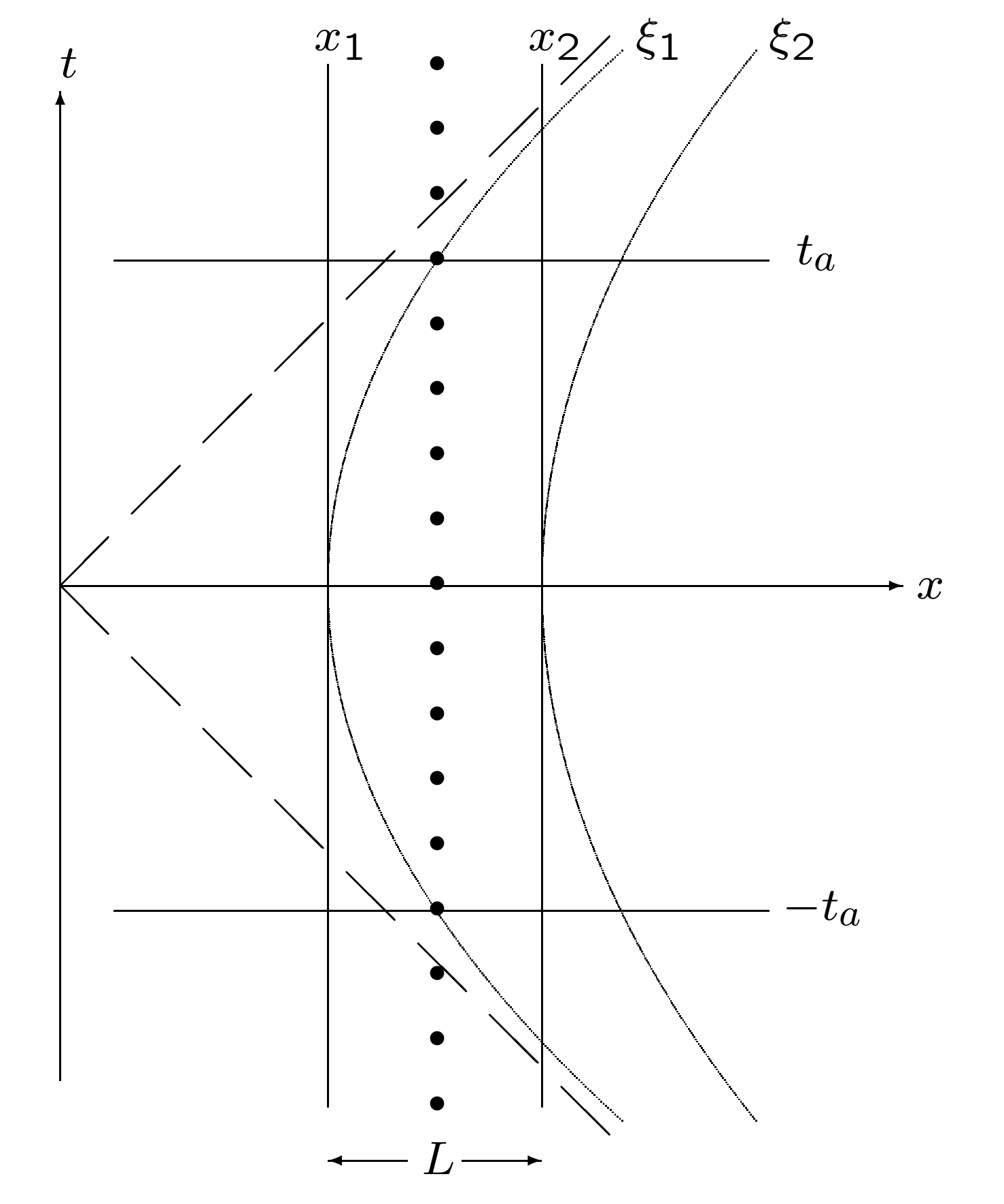}
\end{center}
\caption{Alice and Rob's mirrors align as Rob's cavity comes to rest at $t=0$. The vertical lines labelled $x_1$ and $x_2$ are Alice's mirrors, the curved lines labelled $\xi_1$ and $\xi_2$ are Rob's accelerating mirrors. The dotted line is the atom trajectory with short horizontal lines labelling $\pm t_a$ where the atom moves in and out of alignment with Rob's cavity. The dashed line represents the causal horizon in Rindler coordinates.}
\end{figure}

The dynamics of the  field inside the cavity is given by the Klein-Gordon equation,
$g^{\mu\nu}\nabla_{\mu}\nabla_{\nu}\phi=0$, 
where $g^{\mu\nu}$ is the metric tensor and $\nabla_{\mu}$ is a covariant derivative. 
The perfectly reflecting mirrors impose Dirichlet boundary conditions on the field which is set to vanish on the boundary. The solutions to the equation are given by plane waves,
\eqn{
u_{n}(t,x)\eq \frac{1}{\sqrt{n\pi}}\sin\left(\frac{n\pi}{L}[x-x_1]\right)e^{-\frac{in\pi}{L}t},
}
and the quantum field contained within the cavity walls is $
\hat{\phi}_A(t,x)=\sum_n(u_{n}(t,x)\hat{a}_n+u_{n}^*(t,x)\hat{a}_n^{\dagger})$
where $\hat{a}_n^{\dagger}$ and $\hat{a}_n$ are creation and annihilation operators with $[\hat{a}_n,\hat{a}^{\dagger}_{n^{\prime}}]=\delta_{n,n^{\prime}}$. The Minkowski vacuum state is defined by $\hat{a}_n|0\rangle_A=0$ $\forall n$ where the subscript $A$ indicates that these are states in Alice's inertial cavity. We also consider an observer  moving with uniform acceleration named Rob. Rob is also in possession of a cavity which, in this case, is described by uniformly accelerating boundary conditions. Suitable coordinates $(\eta,\xi)$ for uniform acceleration are called Rindler coordinates and are defined by the transformation,
\begin{equation}
t= a^{-1}e^{a\xi}\mathrm{sinh}(a\eta), \qquad
x= a^{-1}e^{a\xi}\mathrm{cosh}(a\eta).
\end{equation}
The coordinates cover only a portion of Minkowski space-time known as the right Rindler wedge which is bounded by a causal horizon. If we consider Rob to be stationary in Rindler coordinates with constant spatial location $\xi=\xi_1$ for all $\eta$, his trajectory in Minkowski coordinates is given by $x_1(t)=\sqrt{t^2+X_1^2}$, where $X_1=a^{-1}e^{a\xi_1}$. Furthermore, Rob's proper acceleration is given by $\alpha=X_1^{-1}$. Without loss of generality, we will choose $\xi_1=0$ which will result in the proper acceleration being equal to the parameter $a$. 

The uniformly accelerating cavity will consist of two mirrors, stationary with respect to Rob, one at $\xi_1$ and the other at $\xi_2$. We choose Alice and Rob to meet at $t=0$ with their mirrors aligned as shown in Fig(1). This fixes the position of Alice's cavity to be $x_1=X_1$. It also fixes the length of Rob's cavity at $t=0$ to be $X_2-X_1=L$.  The length of Rob's cavity in Rindler coordinates will therefore be $L^{\prime}=\frac{1}{a}\ln\left(1+aL\right)$ for all $t$, which is a constant for fixed values of the kinematical parameter $a$.

We now need to solve the Klein-Gordon equation with the uniformly accelerated boundary conditions described above. This equation can be solved in either Minkowski or Rindler coordinates. Solving the Klein-Gordon equation between Rob's mirrors using Minkowski coordinates would allow us to describe particles which are positive energy excitations with respect to Minkowski time, that is we would be able to describe the particle content in the cavity as seen by an inertial observer moving through the cavity. However, we are interested in describing the particle content in the cavity as seen by Rob, who moves along the cavity's trajectory making measurements and other quantum operations on the field.  In this case, the boundary conditions $\phi[\eta,\xi_1]=\phi[\eta,\xi_2]=0$ are time-independent since the length $L^{\prime}$ is a constant. The equation can be solved immediately due to its conformal invariance. The solutions take the usual form,
\eqn{
v_{n}(\eta,\xi)\eq \frac{1}{\sqrt{n\pi}}\sin\left(\frac{n\pi}{L^{\prime}}\xi\right)e^{-\frac{in\pi}{L^{\prime}}\eta},
} 
where $n=1,2,\dots$ Therefore, the quantum field inside the cavity from Rob's perspective is given by $\hat{\phi}_R(\eta,\xi)=\sum_n(v_n(\eta,\xi)\hat{b}_n+v_n^*(\eta,\xi)\hat{b}_n^{\dagger})$ where $\hat{b}_n^{\dagger}$ and $\hat{b}_n$ are once more creation and annihilation operators with $[\hat{b}_n,\hat{b}^{\dagger}_{n^{\prime}}]=\delta_{n,n^{\prime}}$. The ground state, in this case, is defined by $\hat{b}_n|0\rangle_R=0$, $\forall n$ where the subscript $R$ indicates that these are states in a Rindler cavity. Note that the Rindler coordinates completely cover the region inside the cavity and that the horizon in Rindler coordinates always lies outside the cavity for all values of $a$. We assume the cavity's mirrors to be perfectly reflecting therefore, if Rob prepares his cavity in a given Rindler state, it will remain in such state for all times \cite{Avagyan:2002p420}. This is in agreement with Schutzhold and Unruh's \cite{Schutzhold:2005p311} comment  concerning how the cavity protects the state from Unruh radiation.

Before we introduce our proposal to entangle the field modes within the inertial and non-inertial cavities, we would like to mention that using cavities has the clear advantage that once the state has been produced, there is no degradation of entanglement due to the Unruh effect. This is in stark contrast with recent proposals of encoding quantum information in the states of  two atoms, one of them being non-inertial \cite{lin:2008, landulfo:2009}, as well as free fields \cite{fuentes:2005,  adesso:2007, pan:2008} and the cavities of Alsing and Milburn \cite{alsing:2003}.  Assume that we are able to prepare a maximally entangled state between the two cavities such as,
\eqnn{
|\Phi\rangle\eq\frac{1}{\sqrt{2}}\left(|0\rangle_A|1_n\rangle_R+|1_m\rangle_A|0\rangle_R\right).
}
where $|1_n\rangle_R=\hat{b}^{\dagger}_n|0\rangle_R$ is a single Rindler particle state in Rob's cavity with frequency $\omega=n\pi/L^{\prime}$ and similarly, the state $|1_m\rangle_A$ is a single Minkowski particle state in Alice's cavity with frequency $\omega=m\pi/L$. The time evolution of this joint state can be calculated using the Hamiltonians governing the fields in each cavity. If we calculate the Von-Neumann entropy defined as $\mathcal{E}(\rho_r)=-\mathrm{Tr}[\rho_r\log(\rho_r)]$ where $\rho_r$ is the reduced density matrix of either Alice or Rob, we find that the entanglement of this state is maximal at all times. Most importantly, the portion of the entangled state inside Rob's cavity is localised away from the Unruh horizon and is defined with respect to Rob's notion of particle, hence he always has full access to the information contained in it. Once the state has been prepared, Alice and Rob will be able to exploit the full entanglement in the state. The key difference between this and the free field situation is that instead of having entanglement between two Minkowski modes, here we have entanglement between a Rindler and a Minkowski mode. In our scenario, it seems natural for Rob to prepare his cavity in a Rindler state while Alice prepares her cavity in a Minkowski state. In the free field situation this would be infeasible.

We now introduce our scheme for entangling the Minkowski field modes contained in the inertial cavity and the Rindler modes contained in the non-inertial cavity, demonstrating how our ability to do this depends on Rob's proper acceleration. A common proposal for producing entangled states between the field modes of two cavities involves the interaction with a single atom moving through the cavities. Such proposals have already been experimentally realised \cite{Raimond:2001p6}. Unfortunately, in most cases, these schemes are extremely sensitive to small variations in the parameters of the interaction. However, Browne and Plenio \cite{Browne:2003p374} have proposed a non-deterministic scheme which is robust for a wide range of effective interaction times. The setup is as follows; a two-level atom is prepared in its excited state and  passed through two cavities which are prepared in their ground states. The atom is subsequently measured. If the atom is found in its ground state, the atom must have emitted a photon into one of the cavities and hence, they are  in an entangled superposition. In order to make this protocol near-deterministic, Alice and Bob can attempt it simultaneously on many systems, keeping only those that succeed. Here we generalise this scenario to the situation where one of the cavities is uniformly accelerating. Fig.(2) shows the situation at $t=0$ when the two cavities are aligned. 
\begin{figure}[h] \label{fig:figure2}
	\includegraphics[width=6cm]{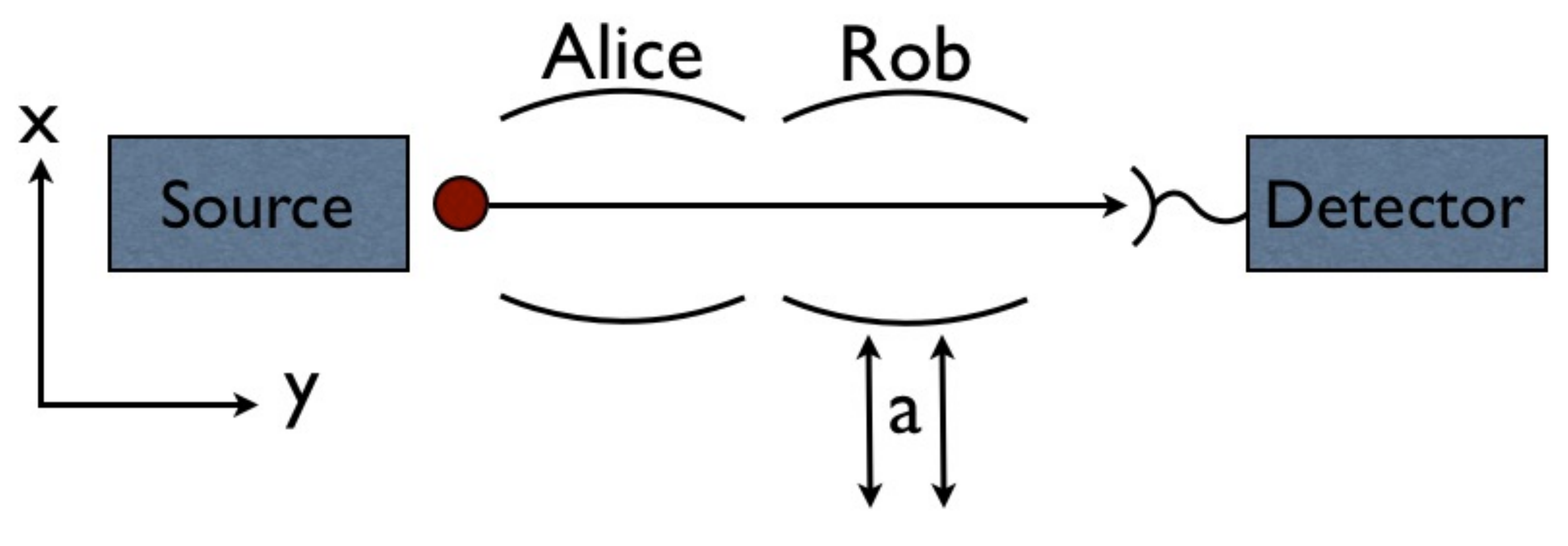} 
	\caption{Entangling two cavities}
\end{figure}
The interaction between the atom and the field is modelled using the Unruh-DeWitt detector \cite{crispino:2008}. Initially the atom is in its excited state and the cavities in the ground state with respect to the coordinate system of their corresponding owner. After the interaction and subsequent projection onto the ground state of the atom, the state of the field, in first order perturbation theory is given by \cite{birrell:1982},
\begin{eqnarray}
|\Phi\rangle&=&-i\int  m(\tau)\left(\epsilon_A(\tau)\hat{\phi}_A[x(\tau)]+\epsilon_R(\tau)\hat{\phi}_R[x(\tau)]\right)d\tau|0\rangle \nonumber 
\end{eqnarray}
were $m(\tau)=\left(\sigma^+e^{-i\Omega\tau}+\mathrm{h.c.}\right)$ is the monopole operator for the detector and $|0\rangle=|0\rangle_A|0\rangle_R$. The functions $\epsilon_{A}(\tau)$ and $\epsilon_{R}(\tau)$ model the effective interaction time as the atom passes through the length of Alice's and Rob's cavity in the y-direction. The $\sigma^{\pm}$ are raising and lowering operators for the atom which has an energy level spacing given by $\hbar\Omega$. Here $\tau$ is the proper time for the atom and $x(\tau)$ are the coordinates of its trajectory parameterised by its proper time. We consider an atom trajectory which will pass straight through the centre of the stationary cavity, passing through the centre-point of the accelerating cavity as the cavities become aligned at $t=0$. Note that the atom interacts with the value of the field along its trajectory therefore, the interaction will change as the atom moves closer to the walls of the accelerating cavity. In Minkowski coordinates the atom's  trajectory is given by $x(\tau)=(t(\tau),x(\tau))=(\tau,X)$ where $X=x_1+L/2$ is the spatial location of the atom. From the Rindler perspective the trajectory is,
\eqn{
(\eta(\tau),\xi(\tau))=\left(\frac{1}{a}\mathrm{artanh}\left(\frac{\tau}{X}\right),\frac{1}{a}\ln\left(a\sqrt{X^2-\tau^2}\right)\right).
}
Note we have suppressed the motion of the atom in the $y$-direction as this is taken into account by the switching function $\epsilon(\tau)$.

The final state of the cavities can be expressed as,
\begin{eqnarray}
|\Phi\rangle&=&\sum_n\left(I_A(n)\hat{a}^{\dagger}_{n}+I_R(n)\hat{b}^{\dagger}_{n}\right)|0\rangle_A|0\rangle_R\nonumber
\end{eqnarray}
For spherical mirrors commonly used in experiments, the modes in the $y$-direction are effectively Gaussian. Therefore, the effective coupling function will be modelled by a Gaussian-like function $\epsilon_A(\tau)=\epsilon e^{-\frac{(\tau-t_A)^2}{W^2}}$, where $W$ depends on both the cavity geometry and the atom velocity in the $y$-direction. The atom passes through the centre of Alice's cavity at $\tau=t_A$ and we obtain the closed form solution,
\eqn{
I_A(n)\eq-\frac{i\epsilon W}{\sqrt{n}}\sin\left(\frac{n\pi}{2}\right)\exp\left[-\frac{\Delta^2W^2}{4}+i\Delta t_A\right]
}
where $\Delta=\frac{n\pi}{L}-\Omega$. This is the probability amplitude for finding a particle in Alice's cavity with frequency $n\pi/L$.  We observe that it is maximum for a particle in resonance with the atom frequency $\Omega$ and it decays exponentially for off resonant frequencies. 

We use the same Gaussian switching function for Rob's cavity but with two modifications. The Gaussian is centred on $\tau=0$ and goes to zero when the cavity becomes completely unaligned with the atom at $\pm t_a$, see Fig(2). To achieve this we set $\epsilon_R(\tau)=\epsilon\exp[-\frac{\left(t_a\mathrm{artanh}[\tau/t_a]\right)^2}{W^2}]$, where $t_a=\sqrt{L/a+\frac{1}{4}L^2}$. Note that in the limit $a\rightarrow0$, $\epsilon_R(\tau)\rightarrow \epsilon e^{-\frac{\tau^2}{W^2}}$. With these considerations we find that,
\begin{eqnarray}
I_R(n)&=&-\frac{i\epsilon}{\sqrt{n\pi}}\int_{-t_a}^{t_a}d\tau\sin\left(\frac{n\pi}{L^{\prime}}\xi(\tau)\right)\times\nonumber \\
&\exp&\left[-\frac{\left(t_a\mathrm{artanh}[\frac{\tau}{t_a}]\right)^2}{W^2}+\frac{in\pi}{L^{\prime}}\eta(\tau)-i\Omega \tau\right].\quad
\end{eqnarray}
Note that in the limit $a\rightarrow0$ we obtain $I_R(n)\rightarrow I_A(n)e^{i\Delta t_A}$. This implies the probability of finding the atom in either Rob's cavity or Alice's cavity will be equal, hence the state produced will be maximally entangled. In contrast to this, in the limit $a\rightarrow\infty$, we obtain $I_R(n)\rightarrow0$ where the probability of finding the atom in Rob's cavity vanishes and the state prepared is completely separable.
We evaluate the integral $I_R(n)$ numerically and then estimate the Von-Neumann entropy of the state after the interaction. We choose $L$ to be on the order of $10\mathrm{cm}$ and $W$ on the order of $10^{-7}\mathrm{s}$. We find that for $a=0$ the state is maximally entangled and the entanglement decreases with the acceleration, vanishing in the limit $a\rightarrow\infty$ as seen in Fig.(3).

\begin{figure}[h]
\includegraphics[width=9cm]{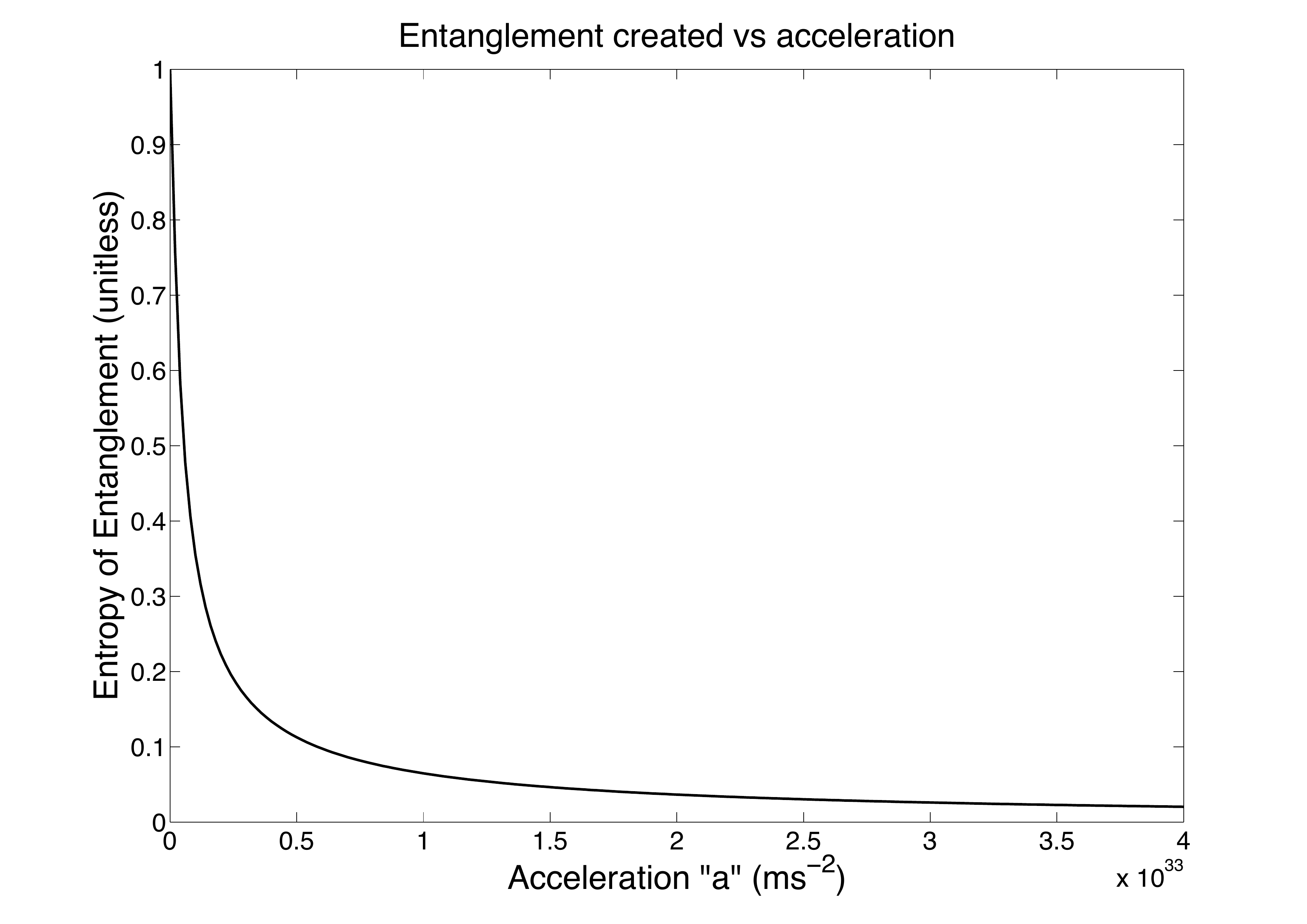} 
\caption{Entanglement as a function of Rob's acceleration}
\end{figure}

Note that, interestingly, the dependence of entanglement with acceleration is qualitatively similar to the degradation of entanglement for free fields due to the Unruh effect \cite{alsing:2003}. However, in our analysis, the entanglement decreases not because of the loss of information in the Rindler horizon but because our ability to entangle the cavities is reduced by the cavity's acceleration. It appears that as acceleration grows the modes inside the cavity become increasingly detuned from the atom reducing it's probability of emission. However, Alice and Rob can correct for this effect by measuring Rob's acceleration and adjusting the parameter $L$ accordingly. As the length $L$ is changed Alice's cavity is brought off resonance while Rob's cavity is brought back closer to resonance. In Fig(4) we consider $a=8\times10^{33} \mathrm{ms}^{-2}$ and show that as $L$ increases the entanglement produced rises up to unity. This shows that it is always possible to create and store a maximally entangled state for any finite acceleration, however the probability of success will decrease with increasing acceleration \footnote{In order to measure the entanglement the accelerating observer must cool his apparatus to its ground-state in a larger device before making the measurements.}.

\begin{figure}[h]
\includegraphics[width=9cm]{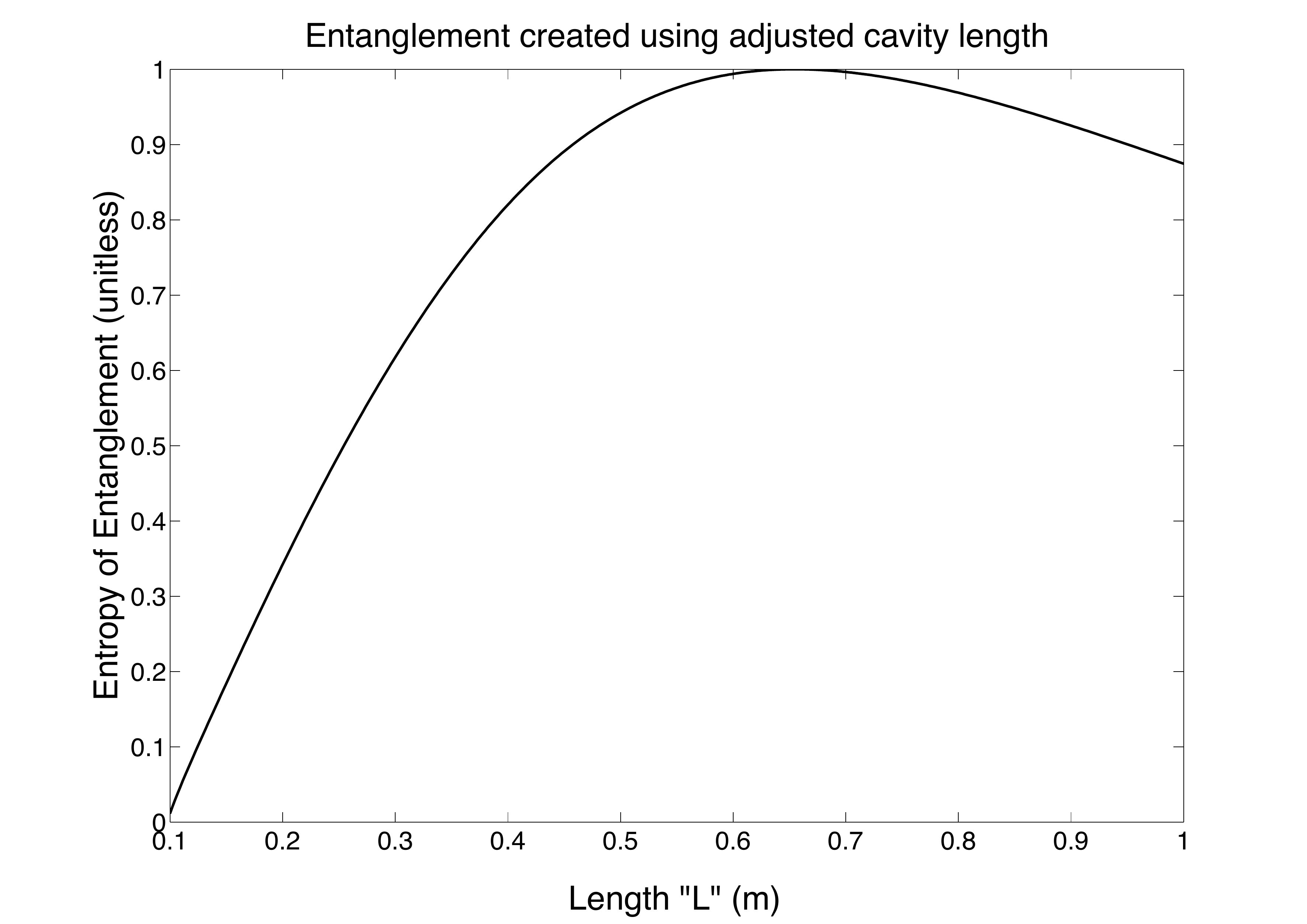} 
\caption{Entanglement as a function of $L$ for $a=8\times10^{33}ms^{-2}$}
\end{figure}

Here we have considered cavities with perfect mirrors. If the non-inertial cavity is considered leaky, some Unruh radiation would couple to the field inside the cavity. We have also restricted to the scalar field. In order to consider the possibility of effects from polarisation the monopole atom we considered would need to be replaced by a full dipole. In this idealised limit of perfect cavities and scalar field we have shown that there is no bound to the entanglement which can be shared between inertial and non-inertial observers. We thank G. Milburn, R. Mann, J. Louko, A. Dragan, E. Martin-Martinez, P. Alsing and B. L. Hu for helpful discussions.


\begin{thebibliography}{16}%
\makeatletter
\providecommand \@ifxundefined [1]{%
 \@ifx{#1\undefined}
}%
\providecommand \@ifnum [1]{%
 \ifnum #1\expandafter \@firstoftwo
 \else \expandafter \@secondoftwo
 \fi
}%
\providecommand \@ifx [1]{%
 \ifx #1\expandafter \@firstoftwo
 \else \expandafter \@secondoftwo
 \fi
}%
\providecommand \natexlab [1]{#1}%
\providecommand \enquote  [1]{``#1''}%
\providecommand \bibnamefont  [1]{#1}%
\providecommand \bibfnamefont [1]{#1}%
\providecommand \citenamefont [1]{#1}%
\providecommand \href@noop [0]{\@secondoftwo}%
\providecommand \href [0]{\begingroup \@sanitize@url \@href}%
\providecommand \@href[1]{\@@startlink{#1}\@@href}%
\providecommand \@@href[1]{\endgroup#1\@@endlink}%
\providecommand \@sanitize@url [0]{\catcode `\\12\catcode `\$12\catcode
  `\&12\catcode `\#12\catcode `\^12\catcode `\_12\catcode `\%12\relax}%
\providecommand \@@startlink[1]{}%
\providecommand \@@endlink[0]{}%
\providecommand \url  [0]{\begingroup\@sanitize@url \@url }%
\providecommand \@url [1]{\endgroup\@href {#1}{\urlprefix }}%
\providecommand \urlprefix  [0]{URL }%
\providecommand \Eprint [0]{\href }%
\@ifxundefined \urlstyle {%
  \providecommand \doi  [0]{\begingroup \@sanitize@url \@doi}%
  \providecommand \@doi [1]{\endgroup \@@startlink {\doibase
  #1}doi:\discretionary {}{}{}#1\@@endlink }%
}{%
  \providecommand \doi  [0]{doi:\discretionary{}{}{}\begingroup
  \urlstyle{rm}\Url }%
}%
\providecommand \doibase [0]{http://dx.doi.org/}%
\providecommand \Doi [0]{\begingroup \@sanitize@url \@Doi }%
\providecommand \@Doi  [1]{\endgroup\@@startlink{\doibase#1}\@@Doi}%
\providecommand \@@Doi [1]{#1\@@endlink}%
\providecommand \selectlanguage [0]{\@gobble}%
\providecommand \bibinfo  [0]{\@secondoftwo}%
\providecommand \bibfield  [0]{\@secondoftwo}%
\providecommand \translation [1]{[#1]}%
\providecommand \BibitemOpen [0]{}%
\providecommand \bibitemStop [0]{}%
\providecommand \bibitemNoStop [0]{.\EOS\space}%
\providecommand \EOS [0]{\spacefactor3000\relax}%
\providecommand \BibitemShut  [1]{\csname bibitem#1\endcsname}%
\bibitem [{\citenamefont {Peres}\ and\ \citenamefont
  {Terno}(2004)}]{peres:2004}%
  \BibitemOpen
  \bibfield  {author} {\bibinfo {author} {\bibfnamefont {A.}~\bibnamefont
  {Peres}}\ and\ \bibinfo {author} {\bibfnamefont {D.~R.}\ \bibnamefont
  {Terno}},\ }\href@noop {} {\bibfield  {journal} {\bibinfo  {journal} {Rev.
  Mod. Phys.},\ }\textbf {\bibinfo {volume} {76}},\ \bibinfo {pages} {93}
  (\bibinfo {year} {2004})}\BibitemShut {NoStop}%
\bibitem [{\citenamefont {Nielsen}\ and\ \citenamefont
  {Chuang}(2000)}]{nielsen:2000}%
  \BibitemOpen
  \bibfield  {author} {\bibinfo {author} {\bibfnamefont {M.}~\bibnamefont
  {Nielsen}}\ and\ \bibinfo {author} {\bibfnamefont {I.}~\bibnamefont
  {Chuang}},\ }\href@noop {} {\emph {\bibinfo {title} {Quantum computation and
  quantum information}}}\ (\bibinfo  {publisher} {CUP},\ \bibinfo {year}
  {2000})\BibitemShut {NoStop}%
\bibitem [{\citenamefont {Fuentes-Schuller}\ and\ \citenamefont
  {Mann}(2005)}]{fuentes:2005}%
  \BibitemOpen
  \bibfield  {author} {\bibinfo {author} {\bibfnamefont {I.}~\bibnamefont
  {Fuentes-Schuller}}\ and\ \bibinfo {author} {\bibfnamefont {R.~B.}\
  \bibnamefont {Mann}},\ }\href@noop {} {\bibfield  {journal} {\bibinfo
  {journal} {Phys. Rev. Lett.},\ }\textbf {\bibinfo {volume} {95}},\ \bibinfo
  {pages} {120404} (\bibinfo {year} {2005})}\BibitemShut {NoStop}%
\bibitem [{\citenamefont {Adesso}\ \emph {et~al.}(2007)\citenamefont {Adesso},
  \citenamefont {Fuentes-Schuller},\ and\ \citenamefont
  {Ericsson}}]{adesso:2007}%
  \BibitemOpen
  \bibfield  {author} {\bibinfo {author} {\bibfnamefont {G.}~\bibnamefont
  {Adesso}}, \bibinfo {author} {\bibfnamefont {I.}~\bibnamefont
  {Fuentes-Schuller}}, \ and\ \bibinfo {author} {\bibfnamefont
  {M.}~\bibnamefont {Ericsson}},\ }\Doi {10.1103/PhysRevA.76.062112} {\bibfield
   {journal} {\bibinfo  {journal} {Phys. Rev. A},\ }\textbf {\bibinfo {volume}
  {76}},\ \bibinfo {pages} {062112} (\bibinfo {year} {2007})}\BibitemShut
  {NoStop}%
\bibitem [{\citenamefont {Pan}\ and\ \citenamefont {Jing}(2008)}]{pan:2008}%
  \BibitemOpen
  \bibfield  {author} {\bibinfo {author} {\bibfnamefont {Q.}~\bibnamefont
  {Pan}}\ and\ \bibinfo {author} {\bibfnamefont {J.}~\bibnamefont {Jing}},\
  }\Doi {10.1103/PhysRevA.77.024302} {\bibfield  {journal} {\bibinfo  {journal}
  {Phys. Rev. A},\ }\textbf {\bibinfo {volume} {77}},\ \bibinfo {pages}
  {024302} (\bibinfo {year} {2008})}\BibitemShut {NoStop}%
\bibitem [{\citenamefont {Lin}\ \emph {et~al.}(2008)\citenamefont {Lin},
  \citenamefont {Chou},\ and\ \citenamefont {Hu}}]{lin:2008}%
  \BibitemOpen
  \bibfield  {author} {\bibinfo {author} {\bibfnamefont {S.-Y.}\ \bibnamefont
  {Lin}}, \bibinfo {author} {\bibfnamefont {C.-H.}\ \bibnamefont {Chou}}, \
  and\ \bibinfo {author} {\bibfnamefont {B.~L.}\ \bibnamefont {Hu}},\ }\Doi
  {10.1103/PhysRevD.78.125025} {\bibfield  {journal} {\bibinfo  {journal}
  {Phys. Rev. D},\ }\textbf {\bibinfo {volume} {78}},\ \bibinfo {pages}
  {125025} (\bibinfo {year} {2008})}\BibitemShut {NoStop}%
\bibitem [{\citenamefont {Landulfo}\ and\ \citenamefont
  {Matsas}(2009)}]{landulfo:2009}%
  \BibitemOpen
  \bibfield  {author} {\bibinfo {author} {\bibfnamefont {A.~G.~S.}\
  \bibnamefont {Landulfo}}\ and\ \bibinfo {author} {\bibfnamefont {G.~E.~A.}\
  \bibnamefont {Matsas}},\ }\Doi {10.1103/PhysRevA.80.032315} {\bibfield
  {journal} {\bibinfo  {journal} {Phys. Rev. A},\ }\textbf {\bibinfo {volume}
  {80}},\ \bibinfo {pages} {032315} (\bibinfo {year} {2009})}\BibitemShut
  {NoStop}%
\bibitem [{\citenamefont {Unruh}(1976)}]{unruh:1976}%
  \BibitemOpen
  \bibfield  {author} {\bibinfo {author} {\bibfnamefont {W.~G.}\ \bibnamefont
  {Unruh}},\ }\href@noop {} {\bibfield  {journal} {\bibinfo  {journal} {Phys.
  Rev. D},\ }\textbf {\bibinfo {volume} {14}},\ \bibinfo {pages} {870}
  (\bibinfo {year} {1976})}\BibitemShut {NoStop}%
\bibitem [{\citenamefont {Alsing}\ and\ \citenamefont
  {Milburn}(2003)}]{alsing:2003}%
  \BibitemOpen
  \bibfield  {author} {\bibinfo {author} {\bibfnamefont {P.~M.}\ \bibnamefont
  {Alsing}}\ and\ \bibinfo {author} {\bibfnamefont {G.~J.}\ \bibnamefont
  {Milburn}},\ }\Doi {10.1103/PhysRevLett.91.180404} {\bibfield  {journal}
  {\bibinfo  {journal} {Phys. Rev. Lett.},\ }\textbf {\bibinfo {volume} {91}},\
  \bibinfo {pages} {180404} (\bibinfo {year} {2003})}\BibitemShut {NoStop}%
\bibitem [{\citenamefont {Sch{\"u}tzhold}\ and\ \citenamefont
  {Unruh}(2005)}]{Schutzhold:2005p311}%
  \BibitemOpen
  \bibfield  {author} {\bibinfo {author} {\bibfnamefont {R.}~\bibnamefont
  {Sch{\"u}tzhold}}\ and\ \bibinfo {author} {\bibfnamefont {W.~G.}\
  \bibnamefont {Unruh}},\ }\href {http://arxiv.org/abs/quant-ph/0506028v1}
  {\bibfield  {journal} {\bibinfo  {journal} {arXiv},\ }\textbf {\bibinfo
  {volume} {quant-ph}} (\bibinfo {year} {2005})},\ \Eprint
  {http://arxiv.org/abs/quant-ph/0506028v1} {quant-ph/0506028v1} \BibitemShut
  {NoStop}%
\bibitem [{\citenamefont {Avagyan}\ \emph {et~al.}(2002)\citenamefont
  {Avagyan}, \citenamefont {Saharian},\ and\ \citenamefont
  {Yeranyan}}]{Avagyan:2002p420}%
  \BibitemOpen
  \bibfield  {author} {\bibinfo {author} {\bibfnamefont {R.~M.}\ \bibnamefont
  {Avagyan}}, \bibinfo {author} {\bibfnamefont {A.~A.}\ \bibnamefont
  {Saharian}}, \ and\ \bibinfo {author} {\bibfnamefont {A.~H.}\ \bibnamefont
  {Yeranyan}},\ }\Doi {10.1103/PhysRevD.66.085023} {\bibfield  {journal}
  {\bibinfo  {journal} {Phys. Rev. D},\ }\textbf {\bibinfo {volume} {66}},\
  \bibinfo {pages} {085023} (\bibinfo {year} {2002})}\BibitemShut {NoStop}%
\bibitem [{\citenamefont {Raimond}\ \emph {et~al.}(2001)\citenamefont
  {Raimond}, \citenamefont {Brune},\ and\ \citenamefont
  {Haroche}}]{Raimond:2001p6}%
  \BibitemOpen
  \bibfield  {author} {\bibinfo {author} {\bibfnamefont {J.~M.}\ \bibnamefont
  {Raimond}}, \bibinfo {author} {\bibfnamefont {M.}~\bibnamefont {Brune}}, \
  and\ \bibinfo {author} {\bibfnamefont {S.}~\bibnamefont {Haroche}},\
  }\href@noop {} {\bibfield  {journal} {\bibinfo  {journal} {Rev. Mod. Phys.},\
  }\textbf {\bibinfo {volume} {73}},\ \bibinfo {pages} {565} (\bibinfo {year}
  {2001})}\BibitemShut {NoStop}%
\bibitem [{\citenamefont {Browne}\ and\ \citenamefont
  {Plenio}(2003)}]{Browne:2003p374}%
  \BibitemOpen
  \bibfield  {author} {\bibinfo {author} {\bibfnamefont {D.~E.}\ \bibnamefont
  {Browne}}\ and\ \bibinfo {author} {\bibfnamefont {M.~B.}\ \bibnamefont
  {Plenio}},\ }\Doi {10.1103/PhysRevA.67.012325} {\bibfield  {journal}
  {\bibinfo  {journal} {Phys. Rev. A},\ }\textbf {\bibinfo {volume} {67}},\
  \bibinfo {pages} {012325} (\bibinfo {year} {2003})}\BibitemShut {NoStop}%
\bibitem [{\citenamefont {Crispino}\ \emph {et~al.}(2008)\citenamefont
  {Crispino}, \citenamefont {Higuchi},\ and\ \citenamefont
  {Matsas}}]{crispino:2008}%
  \BibitemOpen
  \bibfield  {author} {\bibinfo {author} {\bibfnamefont {L.~C.~B.}\
  \bibnamefont {Crispino}}, \bibinfo {author} {\bibfnamefont {A.}~\bibnamefont
  {Higuchi}}, \ and\ \bibinfo {author} {\bibfnamefont {G.~E.~A.}\ \bibnamefont
  {Matsas}},\ }\Doi {10.1103/RevModPhys.80.787} {\bibfield  {journal} {\bibinfo
   {journal} {Rev. Mod. Phys.},\ }\textbf {\bibinfo {volume} {80}},\ \bibinfo
  {pages} {787} (\bibinfo {year} {2008})}\BibitemShut {NoStop}%
\bibitem [{\citenamefont {Birrell}\ and\ \citenamefont
  {Davies}(1982)}]{birrell:1982}%
  \BibitemOpen
  \bibfield  {author} {\bibinfo {author} {\bibfnamefont {N.~D.}\ \bibnamefont
  {Birrell}}\ and\ \bibinfo {author} {\bibfnamefont {P.~C.~W.}\ \bibnamefont
  {Davies}},\ }\href@noop {} {\emph {\bibinfo {title} {Quantum fields in curved
  space}}}\ (\bibinfo  {publisher} {CUP},\ \bibinfo {year} {1982})\BibitemShut
  {NoStop}%
\bibitem [{Note1()}]{Note1}%
  \BibitemOpen
  \bibinfo {note} {In order to measure the entanglement the accelerating
  observer must cool his apparatus to its ground-state in a larger device
  before making the measurements.}\BibitemShut {Stop}%
\end{thebibliography}
\end{document}